%% file: main.tex
\documentclass[12pt,titlepage]{article}

\usepackage{float}
\input{macs}

\newif\ifdraft
\drafttrue 

\usepackage[pdftex]{graphicx}
\usepackage{amsmath}
\usepackage{textcomp}
\usepackage{caption}
\DeclareGraphicsExtensions{.jpg,.pdf,.mps,.png}

\usepackage{helvet}
\usepackage[]{amsmath}

\usepackage{cite}
\let\citeleft=(
\let\citeright=)

\graphicspath{{images/}}

\begin{document}
\bibliographystyle{mrm}

\input{titleAbstract}

\input{body}

\newpage
\bibliography{refs_cbaron}

\include{figs}


\end{document}

%% file: macs.tex
\newcommand{\be}{\begin{equation}}
\newcommand{\ee}{\end{equation}}

\newcommand{\bi}{\begin{itemize}}
\newcommand{\ei}{\end{itemize}}

%% file: titleAbstract.tex
\begin{titlepage}

\begin{center}
	\begin{Large}
		\begin{bf}
 High-Resolution, Respiratory-Resolved Coronary MRA Using a Phyllotaxis-Reordered Variable-Density 3D Cones Trajectory
\\ [0.1in]
		\end{bf}
	\end{Large}
\end{center}
\bigskip
\begin{center}
AUTHORS - Srivathsan P. Koundinyan$^1$, Corey A. Baron$^2$, Mario O. Malav\'e$^1$, Frank Ong$^1$, Nii Okai Addy$^1$, Joseph Y. Cheng$^{1, 3}$, Phillip C. Yang$^4$, Bob S. Hu$^{1, 5}$, Dwight G. Nishimura$^1$

\end{center}
\vspace*{0.1in}
\noindent 1. Department of Electrical Engineering,
Stanford University, Stanford, California.  

\noindent 2. Department of Medical Biophysics, 
Western University, London, Ontario.  

\noindent 3.  Department of Radiology, 
Stanford University, Stanford, California. 

\noindent 4. Department of Cardiovascular Medicine, 
Stanford University, Stanford, California.  

\noindent 5. Department of Cardiology, 
Palo Alto Medical Foundation, Palo Alto, California.

\noindent
{\em Running head:} \\
High-Resolution, Respiratory-Resolved Coronary MRA Using a Phyllotaxis-Reordered Variable-Density 3D Cones Trajectory
 
\noindent

\noindent
{\em Address correspondence to:} \\
	Srivathsan P. Koundinyan \\
	Packard Electrical Engineering, Room 355 \\
	350 Serra Mall, Stanford, CA 94305-9510 \\
	Email: skoundin@stanford.edu
    
\noindent
This work was supported by NIH R01 HL127039, NIH T32HL007846, and GE Healthcare.

\noindent
Word Count: 215  (abstract)  3809  (body) 

\end{titlepage}

\section*{Abstract}
\setlength{\parindent}{0in}

Purpose: To develop a respiratory-resolved motion-compensation method for free-breathing, high-resolution coronary magnetic resonance angiography using a 3D cones trajectory. 

Methods: To achieve respiratory-resolved 0.98 mm resolution images in a clinically relevant scan time, we undersample the imaging data with a variable-density 3D cones trajectory. For retrospective motion compensation, translational estimates from 3D image-based navigators (3D iNAVs) are used to bin the imaging data into four phases from end-expiration to end-inspiration. To ensure pseudo-random undersampling within each respiratory phase, we devise a phyllotaxis readout ordering scheme mindful of eddy current artifacts in steady state free precession imaging. Following binning, residual 3D translational motion within each phase is computed using the 3D iNAVs and corrected for in the imaging data. The noise-like aliasing characteristic of the combined phyllotaxis and cones sampling pattern is leveraged in a compressed sensing reconstruction with spatial and temporal regularization to reduce aliasing in each of the respiratory phases. 

Results: In a volunteer and 5 patients, respiratory motion compensation using the proposed method yields improved image quality compared to non-respiratory-resolved approaches with no motion correction and with 3D translational correction. Qualitative assessment by two cardiologists indicates the superior sharpness of coronary segments reconstructed with the proposed method (P \textless{} 0.01). 

Conclusion: The proposed method better mitigates motion artifacts in free-breathing, high-resolution coronary angiography exams compared to translational correction. 


\setlength{\parindent}{0in}
{\bf Key words: coronary angiography, 3D cones, readout ordering, phyllotaxis, retrospective motion correction, respiratory resolved}
\newpage

%% file: body.tex
\section*{Introduction}
Coronary magnetic resonance angiography (CMRA) is a noninvasive technique for evaluating coronary artery disease. Compared to multidetector coronary CT (MDCT), which is the primary alternative noninvasive coronary imaging technique, CMRA presents two key advantages: (1) no ionizing radiation and (2) no artifacts from coronary calcification \cite{rochitte2013computed, hoffmann2006coronary, sakuma2006detection}. MDCT, however, is capable of rapid 3D imaging within a single heartbeat, which minimizes artifacts due to physiological motion. Furthermore, MDCT offers high sub-millimeter spatial resolution for assessment of stenosis. In CMRA, most clinical exams span several minutes while achieving spatial resolution of approximately 1.1 mm \cite{sakuma2006detection, yang2009contrast, kato2010assessment, piccini2014respiratory, prieto2015highly}. 

Recent studies in free-breathing CMRA have attempted to increase spatial resolution and mitigate respiratory motion artifacts with a variety of techniques. Angiograms with an in-plane spatial resolution of 0.35 mm have been demonstrated using diaphragm navigator-gated gradient-echo sequences in concert with parallel imaging \cite{gharib2012feasibility}. By combining compressed sensing with parallel imaging, 1 mm, 0.9 mm, and 0.8 mm isotropic 3D whole-heart coronary imaging techniques have been proposed \cite{pang2015accelerated, akccakaya2014accelerated, addy2015high, bustin2019five}. For motion compensation, these approaches have implemented either retrospective translational or affine correction methods, since diaphragm navigators suffer from low imaging efficiencies that prolong acquisition times \cite{nagel1999optimization, nehrke2001free, moghari2012subject}. However, both translational and affine correction methods can be insufficient to account for the complex nonrigid respiratory motion of the heart \cite{manke2002respiratory, shechter2004respiratory, ingle2014nonrigid, addy20173d, luo2017nonrigid, cruz2017highly, correia2018accelerated}. This limitation is particularly important to address in high-resolution CMRA, where inaccurate motion correction may detract from the benefits of improved spatial resolution.

Extra-dimensional iterative golden-angle sparse parallel (XD-GRASP) imaging has emerged as an effective approach for respiratory motion compensation \cite{feng2016xd}. This approach sorts free-breathing data into different respiratory states to suppress motion artifacts. For CMRA, XD-GRASP has been combined with 3D projection-reconstruction (3DPR) and 3D Cartesian sampling techniques to achieve 1.15 mm isotropic resolution and 1 mm x 1 mm x 2 mm anisotropic resolution, respectively \cite{piccini2017four, feng20185d, correia2018optimized}. 

In this work, we combine the strengths of retrospective translational correction and respiratory-resolved compressed sensing using 3D image-based navigators (3D iNAVs) and a variable-density 3D cones trajectory to demonstrate free-breathing CMRA with 0.98 mm isotropic spatial resolution. The 3D cones trajectory, like 3DPR, offers improved robustness to motion and flow artifacts compared to 3D Cartesian. 3D cones sampling additionally provides the advantage of improved SNR performance and fewer aliasing artifacts compared to 3DPR for equivalent scan times \cite{gurney2006design}. To augment these features of 3D cones imaging and enhance its suitability for XD-GRASP, we present a phyllotaxis readout ordering technique tailored for a balanced steady state free precession (bSSFP) acquisition. For motion compensation beyond XD-GRASP, 3D translational correction within different respiratory phases using 3D iNAVs is also leveraged. Performance of the overall respiratory-resolved technique is compared with global 3D translational correction in volunteer and patient studies for high-resolution CMRA.

\section*{Methods}

\subsection*{Image Acquisition and Trajectory Design}

Within each heartbeat of our free-breathing, cardiac-triggered 3D CMRA sequence (Supporting Information Figure S1), a fat saturation module is followed immediately by imaging data collection during diastole with 18 readouts (99 ms temporal resolution) using a 3D cones $k$-space trajectory \cite{wu2013free}. Alternating-TR bSSFP (ATR bSSFP) is used for further fat suppression and to generate high blood signal. A 3D iNAV of the same imaging region is acquired every heartbeat after collection of the imaging data to monitor beat-to-beat motion throughout the scan. Each 3D iNAV is acquired with a resolution of 4.4 mm in 176 ms using 32 cones readouts \cite{addy20173d}. 

To fully sample a 28x28x14 cm\textsuperscript{3} FOV with 1.2 mm isotropic spatial resolution, our most recent cones-based coronary sequence used 10,980 readouts \cite{malave2019whole}. Increasing the resolution by approximately 20\% to 0.98 mm while maintaining the same FOV would require 16,661 readouts to satisfy the Nyquist criteria. However, to maintain the same scan time, we designed a variable-density version of the 3D cones trajectory \cite{addy2015high}. Here, the sampling density ($f$) of $k$-space ($|\boldsymbol{k}|$) is modified in the following manner: 
\begin{equation} 
f(|\boldsymbol{k}|) = 
\begin{cases} 
      f_1 & |\boldsymbol{k}| \in [0, k_1] \\
      (f_1 - f_2)(1 - (|\boldsymbol{k}| - k_1)/(k_{max} - k_1))^{p} + f_2 &  |\boldsymbol{k}| \in (k_1, k_{max}] \\
   \end{cases}
\end{equation}
where the constant $f_1$ denotes the sampling density in the calibration region spanning from 0 cm\textsuperscript{-1} to $k_1$ cm\textsuperscript{-1}. The transition in sampling density, from $f_1$ at $k_1$ down to $f_2$ at the maximum $k$-space extent ($k_{max}$), is governed by a $p$\textsuperscript{th} order polynomial. We fix $k_1$ = 1020 cm\textsuperscript{-1} (20\% of $k_{max}$), $f_1$ = 1 (corresponding to a fully sampled region), $f_2$ = 0.83, and $p$ = 10. 

\subsection*{Phyllotaxis Readout Ordering}

Our motion correction scheme entails the binning of data into four respiratory phases followed by the application of a compressed sensing reconstruction framework, which benefits from uniform, pseudo-random undersampling patterns \cite{feng2016xd}. Such undersampling patterns can be achieved by leveraging a readout ordering strategy that acquires as distributed a region of $k$-space as possible in each heartbeat \cite{piccini2011spiral}. An added advantage of such a readout ordering strategy is improved robustness to motion artifacts \cite{malave2019whole}. In SSFP-based CMRA, however, artifacts from eddy currents arise with large changes in $k$-space location between successive excitations \cite{bieri2005analysis}. To avoid such artifacts, the standard 3D cones ordering strategy acquires readouts in a clumped, sequential fashion starting from one pole in $k$-space and gradually progressing towards the other pole (Figure 1(A)) \cite{wu2013free}. Phantom and \textit{in vivo} studies acquired with the sequential collection pattern exhibit no apparent eddy current artifacts (Supporting Information Figure S2(A) and Figure 1(B)). However, separating the data into different respiratory phases results in notable gaps in $k$-space when using sequential ordering of readouts (Figure 2(A)).

To spread out the $k$-space data in each heartbeat while reducing eddy current artifacts, we adapt the phyllotaxis ordering approach proposed by Malav\'e, $et$ $al$. for 3D cones imaging \cite{malave2019whole, piccini2011spiral}. In this approach, the total number of readouts is the product of a Fibonacci number (i.e., the number of phyllotaxis segments) and an integer specifying the number of cones per phyllotaxis segment. The set of cone interleaves within a phyllotaxis segment span from the south to the north pole in $k$-space. The number of cones per phyllotaxis segment must be an integer multiple of the number of cones acquired per heartbeat. This avoids large changes in $k$-space location that would arise if the cones jumped to a different phyllotaxis segment during one heartbeat. For a spatial resolution of 1.2 mm, Malav\'e, $et$ $al$. used 610 as the Fibonacci number and 18 as the number of cones per phyllotaxis segment (10,980 total readouts), while collecting 18 cones per heartbeat. This enables traversal from the south to the north pole in $k$-space in a single heartbeat (Figure 1(C)).  

Compared to 1.2 mm imaging demonstrated in our earlier phyllotaxis work \cite{malave2019whole}, a larger volume of $k$-space is sampled to achieve 0.98 mm resolution, resulting in larger changes in $k$-space location between each of the 18 readouts in one cardiac cycle. As a result, phantom as well as \textit{in vivo} studies exhibit eddy current artifacts (Supporting Information Figure S2(B) and Figure 1(D)). 

To reduce the variation in $k$-space location between successive excitations for the desired 0.98 mm resolution, we modify the phyllotaxis readout ordering scheme by reducing the number of phyllotaxis segments (a Fibonacci number) while concurrently increasing the number of cones per phyllotaxis segment (a multiple of the 18 cones acquired per heartbeat). It is possible, for example, to use 377 (the Fibonacci number preceding 610) for the number of phyllotaxis segments and 36 as the number of cones per phyllotaxis segment. Here, collecting 18 cones per heartbeat enables coverage of half (18 cones per heartbeat / 36 cones per phyllotaxis segment) the $k$-space volume in each cardiac cycle. While this strategy reduces the changes between successive excitations, it requires 13,572 readouts, far greater than the target 10,980 readouts. In consideration of scan time constraints, we use a Fibonacci number of 89 and 126 readouts per phyllotaxis segment. This combination of parameters provides a total number of readouts (11,214) that is as close to 10,980 as possible when applying the phyllotaxis reordering strategy. The reordered 0.98 mm variable-density trajectory designed here corresponds to an acceleration factor of 1.5 (16,661/11,214) compared to a fully sampled cones acquisition. 

Note that our particular adapation of phyllotaxis for 0.98 mm imaging traverses one-seventh the extent of a phyllotaxis segment (18 cones per heartbeat / 126 cones per phyllotaxis segment) in a single heartbeat (Figure 1(E) and Supporting Information Video S1). Although readouts in each cardiac cycle are less distributed compared to the original phyllotaxis implementation used for 1.2 mm imaging, the changes between adjacent readouts in a heartbeat are accordingly reduced, leading to no noticeable eddy current artifacts in phantom and \textit{in vivo} studies (Supporting Information Figure S2(C) and Figure 1(F)). Also, compared to the sequential collection pattern, the implemented cones ordering (1) more rapidly changes the regions of $k$-space collected across breathing cycles and (2) within any particular heartbeat, more widely distributes the $k$-space sampling. These two features, in turn, yield respiratory phases with more uniformly distributed $k$-space data exhibiting temporal incoherence (Figure 2(B)), which improves compressed sensing reconstruction \cite{feng2016xd, piccini2017four, zucker2018free}. 

\section*{Motion Compensation}

Translational and affine correction have been the primary approaches to retrospectively address motion corruption in high-resolution CMRA \cite{pang2015accelerated, addy2015high, bustin2019five}. As demonstrated in recent work, these approaches may not completely alleviate respiratory motion artifacts \cite{ingle2014nonrigid, addy20173d, luo2017nonrigid, cruz2017highly, correia2018optimized, correia2018accelerated}. This limitation is particularly important to address in higher resolution imaging, which is more sensitive to motion. In consideration of this, we investigate the following correction approach: 

(1)	A reference 3D iNAV frame is selected. We designate the navigator with the greatest summed similarity, as measured by a mutual information metric, with all other navigators as the reference \cite{luo2017nonrigid}. To determine beat-to-beat 3D translational estimates, rigid-body registration as described in \cite{rohde2004comprehensive} is applied to spatially align each navigator frame to the identified reference frame. \\
(2)	Using the estimated superior-inferior (SI) translation of the heart, we bin the data into four respiratory phases from end-inspiration to end-expiration. The number of phases is consistent with previous efforts demonstrating effective motion mitigation in this manner \cite{feng2016xd, piccini2017four}. Note that each bin is equal-sized; that is, 2803 readouts (11,214 total readouts/4 bins) are assigned to each respiratory phase. In addition, because of the readout ordering scheme implemented for variable-density cones, retrospectively separating the data achieves pseudo-random undersampling in each bin, which is not feasible with the standard sequential ordering strategy (Figure 2).  \\
(3)	Prior work has suggested the potential for residual translational motion within each bin \cite{correia2018optimized}. In consideration of this, rigid-body registration of the 3D iNAVs is applied separately to each bin to determine the intraphase 3D translational motion within each of the different phases (i.e., each bin has its own reference frame determined as in step 2). The $k$-space data for the respiratory phases are independently adjusted with linear phase terms corresponding to the bin-specific 3D translational estimates.  \\
(4)	We apply a parallel imaging and compressed sensing framework that benefits from the noise-like aliasing arising from undersampling a cones trajectory as well as the relatively uniform $k$-space coverage for each respiratory phase. Specifically, we employ the following L\textsubscript{1}-ESPIRiT reconstruction model \cite{uecker2014espirit, feng2016xd}: 
\begin{equation} \label{eq:ommuruga1}
\underset{M}{\arg\min}\left\|FSM - y\right\|_2^2+\lambda_{1}\left\|D_s(M)\right\|_1+\lambda_{2}\left\|D_t(M)\right\|_1
\end{equation}
where $M$ is the respiratory-resolved images (3D spatial + 1D temporal), $y$ is the acquired $k$-space data reformatted into four different respiratory motion states (with each state corrected for intraphase 3D translational motion), $S$ is the coil sensitivity maps, and $F$ is the multichannel non-uniform Fourier transform operator. While the first term enforces data consistency, the second and third terms enforce sparsity in the spatial finite difference ($D_s$) domain and temporal finite difference ($D_t$) domain, respectively. $\lambda_{1}$ and $\lambda_{2}$ are the corresponding regularization parameters. We solve the overall optimization problem using a first-order primal dual algorithm. Regularization parameters $\lambda_{1}$ and $\lambda_{2}$  were empirically selected as $1$x$10^{-3}$ and $5$x$10^{-3}$, respectively (Supporting Information Figure S3). 

Note that we begin with undersampled, variable-density cones data (not a fully sampled acquisition), and retrospectively employ further acceleration by sorting the data into respiratory phases. The final acceleration factor applied to each of the different motion states is 6 (4 bins x 1.5 acceleration factor for all imaging data combined) compared to a fully sampled cones acquisition. 

\section*{Experimental Setup}

The proposed motion correction framework was assessed in a total of six subjects: one non-contrast-enhanced volunteer acquisition, three non-contrast-enhanced patient acquisitions, and two contrast-enhanced patient acquisitions. All patients were suspected of having coronary artery disease. Certain patients did not have contrast administered as it was not necessary for their prescribed clinical protocol. For the contrast-enhanced patient exams, gadobenate dimeglumine was administered and the CMRA scan began approximately 30 to 35 minutes after contrast injection. The delay following contrast administration accommodated the patients' clinical exam. All scans were carried out on a 1.5 T whole-body GE scanner with maximum slew rate of 150 mT/m/ms and maximum gradient amplitude of 40 mT/m. Participants provided informed consent, and the institutional review board approved the complete scan protocol. The studies were performed using an eight-channel cardiac receive coil with cardiac triggering via a peripheral plethysmograph. The variable-density cones trajectory utilized the following imaging parameters: TR\textsubscript{1}/TR\textsubscript{2} = 4.29/1.15 ms; TE = 0.57 ms; flip angle = 70\textdegree; bandwidth = 250 kHz; FOV = 28x28x14 cm\textsuperscript{3}, resolution = 0.98 mm isotropic. 3D iNAVs were acquired every heartbeat with the same FOV and 4.4 mm isotropic resolution. The total scan time across all subjects spanned 623 heartbeats and ranged from 7 to 10 minutes due to variations in heartrate.

The end-expiration image from the respiratory-resolved motion compensation pipeline with intraphase translational correction was compared to no motion correction and to 3D translational correction. Images without any correction and with 3D translational correction were reconstructed using the following modified L\textsubscript{1}-ESPIRiT framework (also solved with a first-order primal dual algorithm): 
\begin{equation} \label{eq:ommuruga2}
\underset{m}{\arg\min}\left\|FSm - y\right\|_2^2+\lambda\left\|D_s(m)\right\|_1
\end{equation}
where $m$ is the desired 3D image, $y$ is the $k$-space data with or without 3D translational correction, and all other terms are as defined in Eq. [2]. The regularization parameter $\lambda$ was set as $1$x$10^{-3}$. 

Two board-certified cardiologists with experience in CMRA performed qualitative analysis of differences in image quality. Thin-slab maximal intensity projection (MIP) reformats of the right coronary artery (RCA) and left coronary artery (LCA) were generated with OsiriX (Pixmeo, Geneva, Switzerland). Compressed sensing reconstructions with no motion correction, translational correction, and the proposed approach were randomized and presented together. The two blinded readers scored the proximal, medial, and distal segments of the RCA and LCA on a five-point scale: 5-Excellent, 4-Good, 3-Moderate, 2-Poor, 1-Non-diagnostic. Paired two-tailed Student's t-tests were applied to determine significance.

\section*{Results}

Reformatted oblique thin-slab MIP images depicting the RCA for the non-contrast-enhanced volunteer study and a contrast-enhanced patient study are shown in Figure 3(A) and 3(B), respectively. In both cases, 3D translational correction partially sharpens the coronary artery as compared to no motion correction, but does not fully correct for the overall observed degradation in image quality. Similar trends are observed in the non-contrast-enhanced patient study (Figure 3(C)), where the extent of motion corruption is more severe relative to the other shown examples. With the proposed correction framework, all segments of the RCA are visualized with enhanced sharpness. Cross-sectional views (inset images) of the lumen of the coronary arteries accentuate the differences in performance of the three correction techniques.

Figure 4(A), 4(B), and 4(C) present the LCA for the non-contrast-enhanced volunteer study, a contrast-enhanced patient study, and a non-contrast-enhanced patient study, respectively. The volunteer study and non-contrast-enhanced patient study exhibit significant blurring without any correction, to the extent that certain regions of the coronary vessels appear absent. Application of 3D translational correction improves the depiction of the LCA, with further improvements from the proposed method apparent in the magnified cross-sectional views. Specifically, boundaries of the lumen in various segments are best portrayed with the proposed motion-correction framework. 

The results of the qualitative reader studies are shown in Figure 5. Without any correction, the average score for the RCA and LCA is 2.39 and 2.01, respectively, across both readers, all segments, and all subjects. Following 3D translational correction, the scores are 2.75 and 2.69. The best scores of 3.64 and 3.61 are seen when utilizing the motion-resolved technique with intraphase 3D translational correction. Observed discrepancies in scores between no motion correction and the proposed method, and between translational correction and the proposed method, are statistically significant with P $<$ 0.01. The correlation coefficients between each reader for the RCA and LCA are 0.74 and 0.69, respectively. The Bland-Altman statistics for the RCA yield a mean difference of 0.70 and a 95\% confidence interval of [\text{-}1.03, 2.43]. For the LCA, the Bland-Altman statistics indicated a mean difference of 0.94 and a 95\% confidence interval of [\text{-}0.60, 2.49].

Supporting Information Figure S4 demonstrates the benefits from intraphase 3D translational correction. In one patient (non-contrast-enhanced), the distal segment of the RCA in the end-expiration image is better visualized following linear phase compensation. Similarly, for another patient (contrast-enhanced), there is notable improvement in the depiction of a LCA branch after accounting for intraphase translational motion. Note that across all volunteer and subject acquisitions, intraphase correction either preserved or enhanced coronary vessel sharpness. 

\section*{Discussion}

We have presented a method for free-breathing, high-resolution whole-heart CMRA. To obtain images with 0.98 mm isotropic resolution in a scan time comparable to our prior efforts with 1.2 mm resolution, we employed a variable-density 3D cones trajectory. Our proposed motion-correction technique begins with the computation of beat-to-beat SI translation of the heart from 3D iNAVs. Using this estimate, we separate the collected $k$-space data into four bins from end-expiration to end-inspiration. In consideration of potential residual translational motion within each respiratory phase, SI, AP, and RL displacements specific to each bin are derived from 3D iNAVs, and the corresponding $k$-space data are adjusted with linear phase terms. Aliasing from the undersampling in each phase is then mitigated with a reconstruction framework using both spatial and temporal finite differences constraints. 

Parallel imaging and compressed sensing reconstruction paradigms benefit from undersampled trajectories with incoherent, noise-like aliasing patterns. Non-Cartesian 3D cones imaging naturally exhibits this feature \cite{wu2013free, ingle2014nonrigid}. To further add to this characteristic, we devised a modified phyllotaxis readout ordering strategy mindful of the target isotropic 0.98 mm resolution as well as the bSSFP acquisition framework. Prior phyllotaxis work in the context of 1.2 mm cones CMRA with bSSFP contrast demonstrated a capability for the cone interleaves to progress from the south pole to the north pole of $k$-space in a single heartbeat with 18 readouts. However, we found that using a similar reordering scheme with our higher resolution trajectory yielded eddy current artifacts. To mitigate these artifacts, our modified approach reduces the jumps in $k$-space position between consecutive excitations. While our modification leads to less distributed $k$-space sampling within any given heartbeat compared to the original phyllotaxis strategy, it nonetheless provides more uniform undersampling within each of the four respiratory phases compared to the sequential acquisition strategy. This is because, relative to sequential collection, the proposed readout ordering technique exhibits more distributed $k$-space coverage within each heartbeat and additionally, traverses from the bottom to the top of the $k$-space volume more quickly across heartbeats. 

We evaluated the effectiveness of the combined 0.98 mm variable-density cones acquisition protocol, readout ordering technique, and respiratory motion correction method on one volunteer and five patients suspected of coronary artery disease. The benefit from the administration of contrast in patients is apparent in Figure 3 and Figure 4. Specifically, as expected, the non-contrast-enhanced acquisitions consistently exhibited a lower signal-to-noise ratio compared to the contrast-enhanced acquisitions. This, however, did not have an apparent impact on the quality of the achieved motion correction outcomes. Across all cases, as per assessment by two cardiologists, the end-expiration image from our proposed approach outperformed 3D translational correction. Note that the mean reader scores for the RCA and LCA are not consistently better for contrast-enhanced acquisitions relative to non-contrast-enhanced acquisitions. This suggests that the variation in reader scores between different patients are primarily due to differences in the degree of motion corruption. 

Prior work for respiratory-resolved CMRA achieved 1.15 mm isotropic resolution and 1 mm x 1 mm x 2 mm anisotropic resolution applying either 3D Cartesian or 3DPR sampling  techniques \cite{piccini2017four, feng20185d, correia2018optimized}. We capitalize on the efficiency of a 3D cones trajectory for respiratory-resolved 0.98 mm CMRA. Unlike the ghosting observed in 3D Cartesian imaging, motion primarily yields blurring artifacts in 3D cones imaging. Phyllotaxis cones further ensures that motion manifests itself as noise-like, diffuse artifacts (which can be mitigated with compressed sensing) instead of coherent streaking \cite{malave2019whole}. Note also that 3D cones exhibits milder undersampling artifacts relative to 3DPR \cite{gurney2006design}. This feature, coupled with the pseudo-random undersampling pattern from phyllotaxis reordering, yields negligible aliasing in each respiratory phase despite the net acceleration factor of 6 introduced first during acquisition and then in the separation of data into different motion states (Supporting Information Figure S5). 

It is important to note that no attempt was made to account for intraphase residual nonrigid motion. With the voxel-by-voxel displacement information offered by the 3D iNAVs, nonrigid motion correction via a generalized matrix description framework \cite{batchelor2005matrix, odille2008generalized} or with autofocusing \cite{ingle2014nonrigid, addy20173d, luo2017nonrigid} could be integrated into the motion-resolved reconstruction pipeline investigated in this study. Further improvements in image quality may be possible with strategies that combine data from different respiratory phases. Supporting Information Figure S5 demonstrates that coronary arteries are visible in other respiratory phases. Thus, it may be possible to combine images from different phases using registration methods for improved SNR \cite{horn1981determining, rueckert1999nonrigid}. 

A notable limitation of our study is computation time. Although reconstructions are performed using graphics processing units (GPUs) with the SigPy environment \cite{sigpy}, total processing time for the proposed correction method is approximately 8.5 hours, which impedes clinical implementation. This can be significantly reduced using recent research on rapid compressed sensing reconstructions with non-Cartesian trajectories \cite{baron2018rapid}. Beyond significant computation time, another limitation is that we do not make comparisons with conventional methods for CMRA scans such as respiratory-gated Cartesian acquisition protocols. In this work, we focused on investigating the performance of a respiratory-resolved reconstruction framework using a non-Cartesian cones trajectory, and the improvements it offers over correcting all the respiratory phases together with 3D translational correction.

\section*{Conclusion}

In this work, we developed an approach for improved robustness to respiration in a free-breathing, 0.98 mm isotropic resolution acquisition. To reduce the scan time for this desired resolution, we employed acceleration via a variable-density cones trajectory. We then presented a phyllotaxis readout ordering strategy for improved $k$-space coverage within each heartbeat compared to the conventional sequential technique. The phyllotaxis collection pattern facilitated a respiratory-resolved motion-correction framework, which was augmented with intraphase 3D translational correction. Evaluations in volunteer and patient studies of the combined acquisition, motion correction, and reconstruction methodology demonstrated improvements over translational correction for high-resolution CMRA. 

%% file: figs.tex
\begin{figure}[h]
  \centering
      \includegraphics[width=\linewidth]{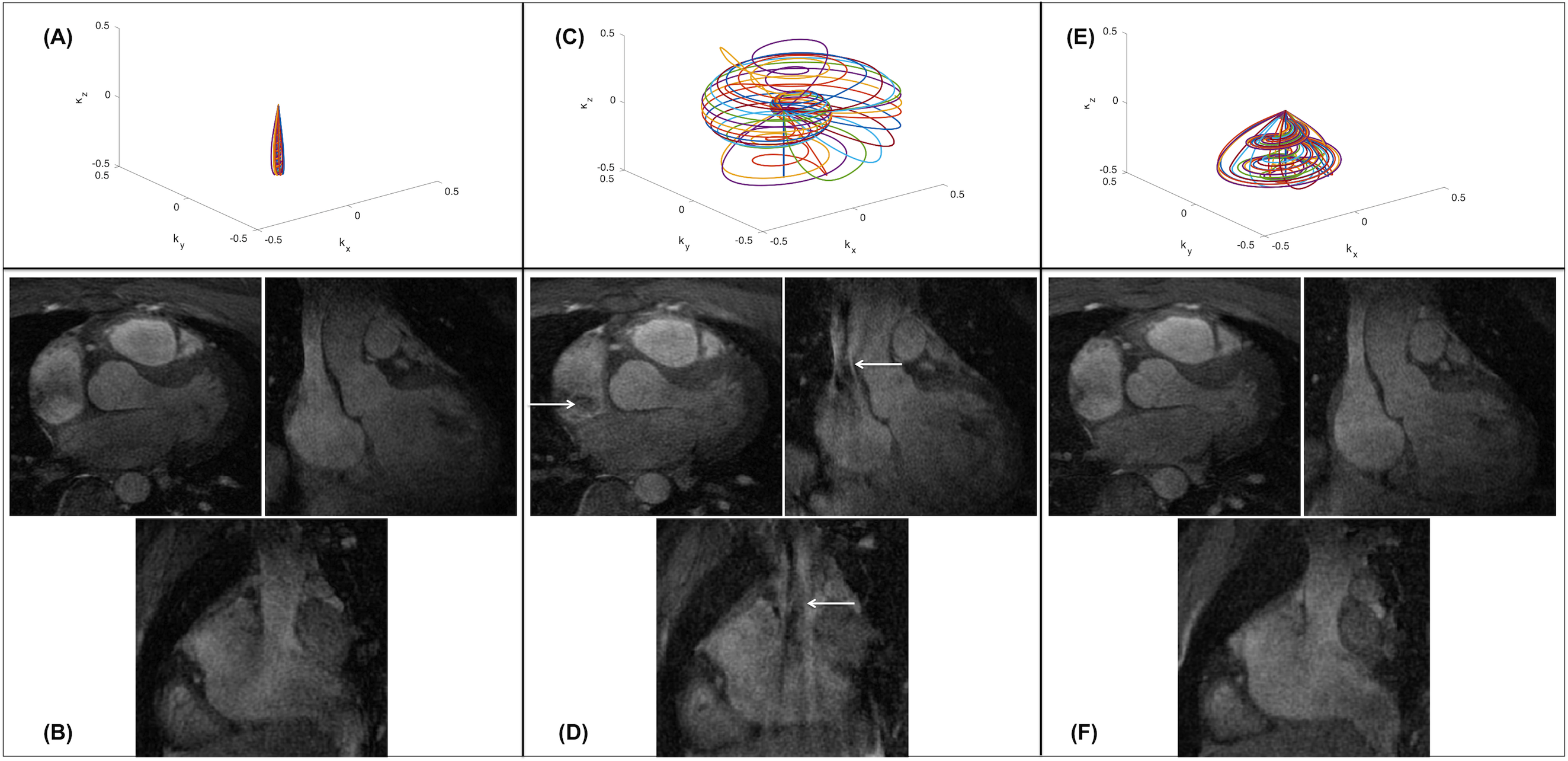}
  \caption[]
    {The standard acquisition in bSSFP cones imaging provides maximal robustness to eddy current artifacts by acquiring readouts in a sequential manner from the south pole to the north pole. When 18 cones per cardiac cycle are acquired, the readouts collected during the first heartbeat for a sequential trajectory are shown in (A). Images obtained in this manner are free of any noticeable artifacts (B). Readouts collected in the first heartbeat when implementing phyllotaxis ordering with a Fibonacci number of 610 and with 18 cones per phyllotaxis segment are displayed in (C). This acquisition scheme results in significant image degradation (white arrows) due to eddy currents (D) for a 0.98 mm isotropic acquisition. Prescribing the Fibonacci number and number of cones per phyllotaxis segment as 89 and 126, respectively, mitigates the changes between consecutive excitations (E), while providing a more distributed coverage compared to a sequential collection pattern. The resulting image exhibits no apparent eddy current artifacts (F). 
    }
\end{figure}

\begin{figure}[h]
  \centering
        \includegraphics[width=\linewidth]{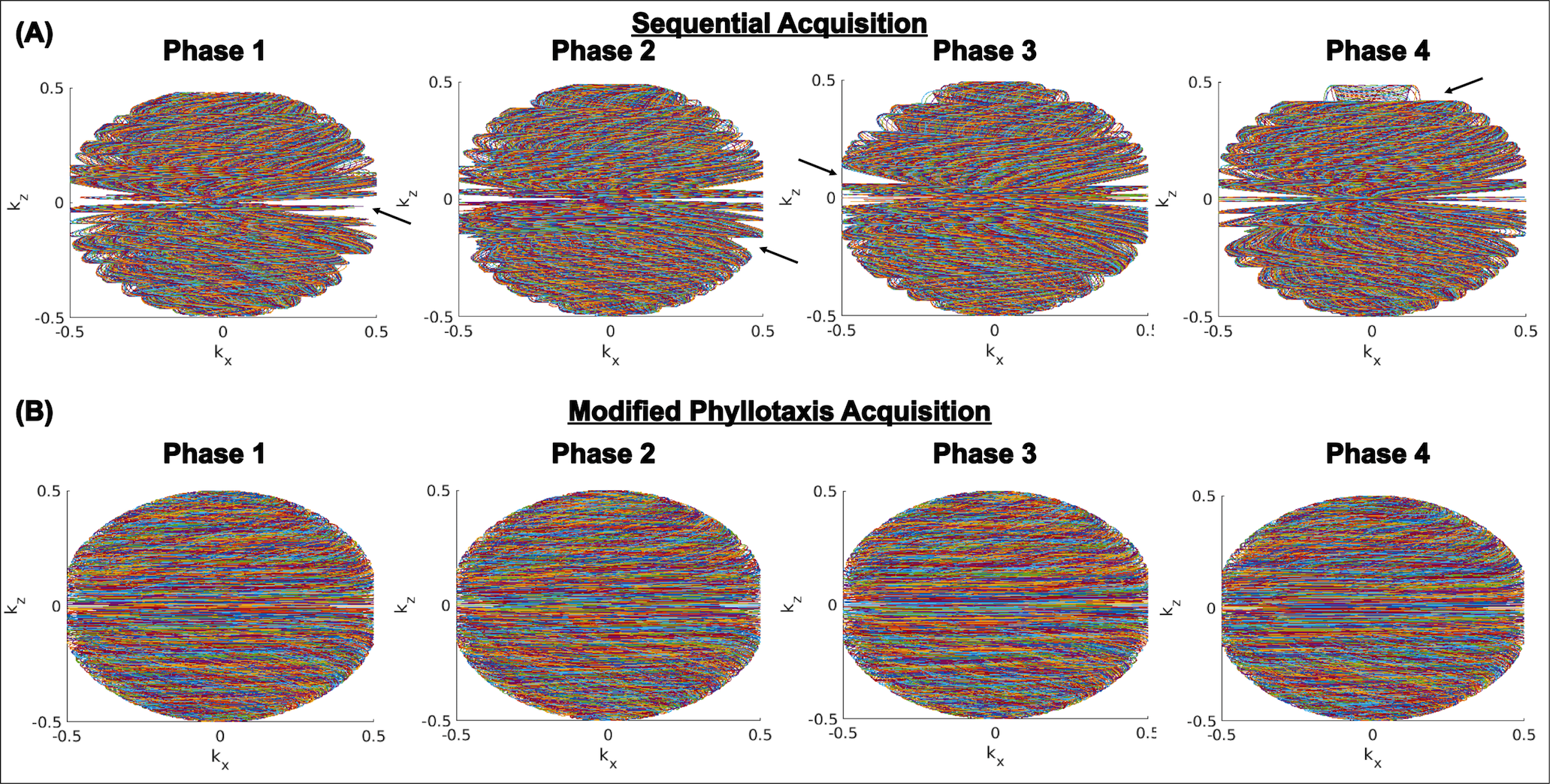}
  \caption[]
    {(A) A sequential acquisition of readouts results in notable gaps (black arrows) in $k$-space when the data is retrospectively binned into 4 respiratory phases. (B) Although the modified phyllotaxis ordering scheme for 0.98 mm isotropic bSSFP CMRA yields less distributed readouts in each heartbeat relative to the original phyllotaxis implementation for 1.2 mm imaging, it nevertheless provides uniform coverage when binned into different respiratory phases. The resulting pseudo-random undersampling in each phase facilitates parallel imaging and compressed sensing reconstruction techniques. 
    }
  \label{fig:theoryConv}
\end{figure}

\begin{figure}[h]
  \centering
        \includegraphics[width=\linewidth]{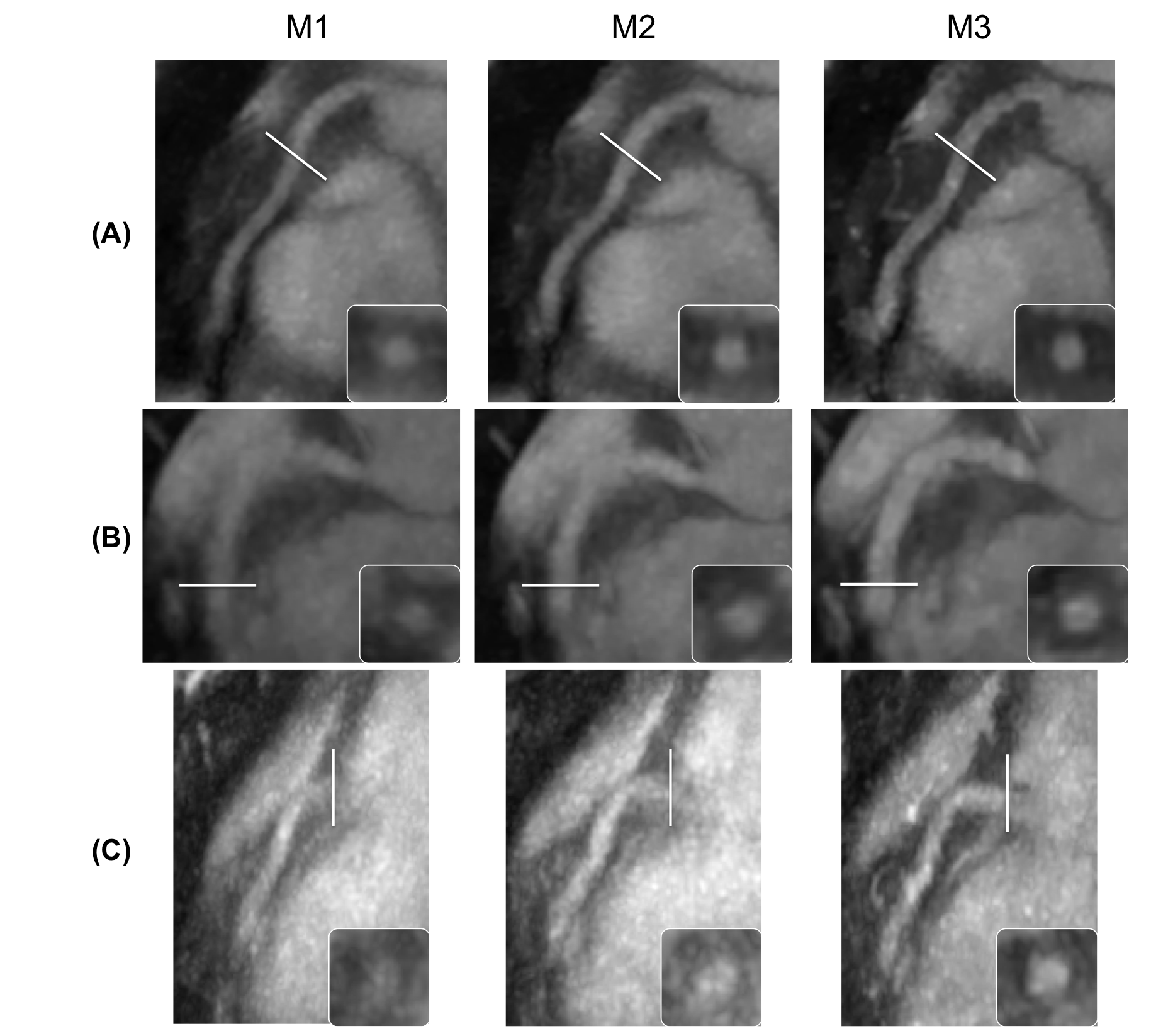}
  \caption[]
    {Reformatted MIP of the RCA following L\textsubscript{1}-ESPIRiT without motion correction (M1) and with translational correction (M2) for one volunteer study (A), one contrast-enhanced patient study (B), and one non-contrast-enhanced patient study (C). The proposed correction framework (M3) yields the sharpest depiction of the RCA compared to the other two approaches. Magnified 2D cross-sectional views of the coronary segments at regions indicated by solid white lines further demonstrate the superior performance of the proposed method. 
    }
\end{figure}

\begin{figure}[h]
  \centering
        \includegraphics[width=\linewidth]{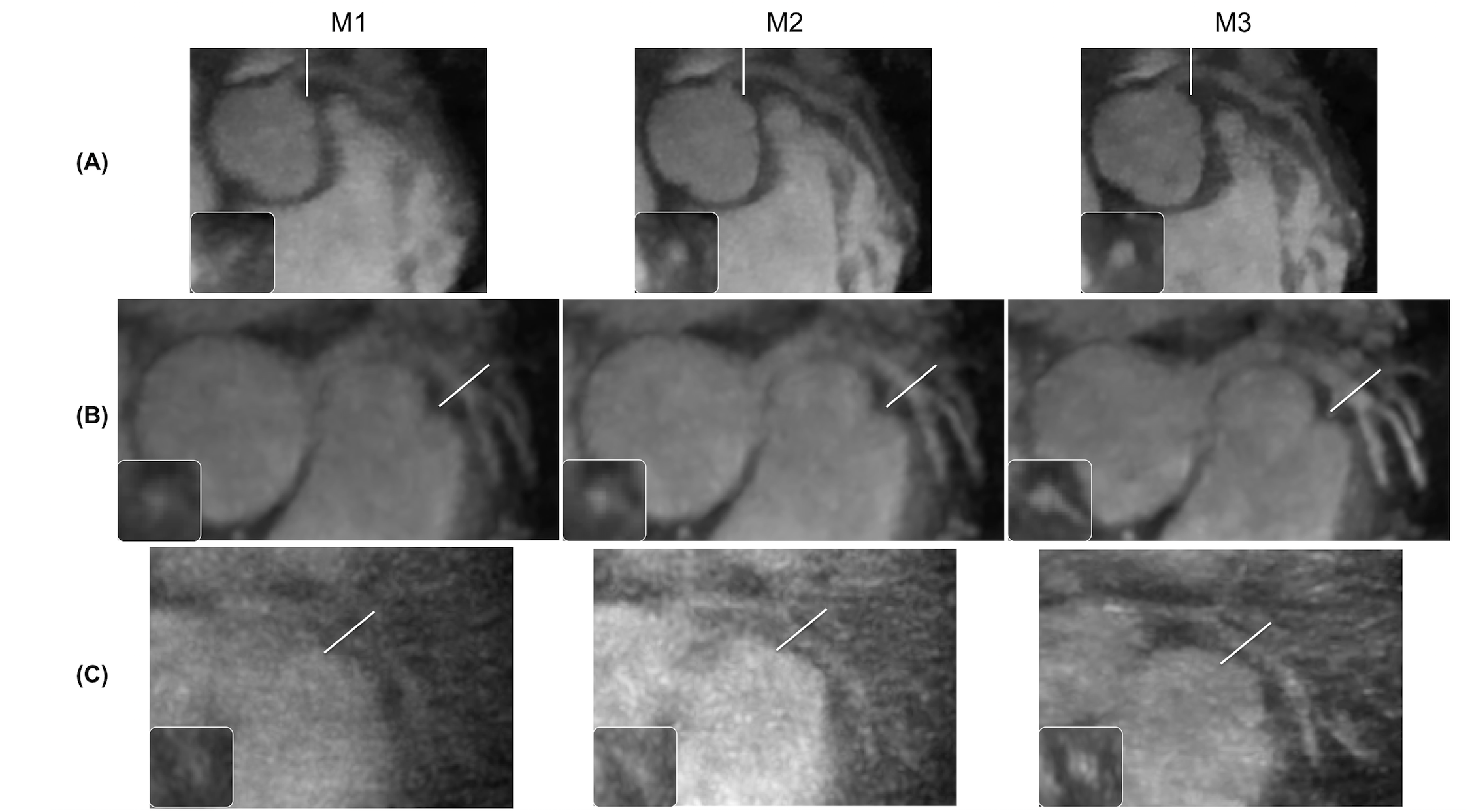}
  \caption[]
    {The LCA without motion correction (M1) and with translational correction (M2) following L\textsubscript{1}-ESPIRiT for one volunteer study (A), one contrast-enhanced patient study (B), and one non-contrast-enhanced patient study (C). As seen with the RCA, the proposed method (M3) presents all regions of the LCA in an enhanced manner relative to M1 and M2. Magnified cross-sectional views of the coronary arteries at segments highlighted by solid white lines show improved depiction of the lumen of the LCA with M3.  
    }
\end{figure}

\begin{figure}[h]
  \centering
        \includegraphics[width=\linewidth]{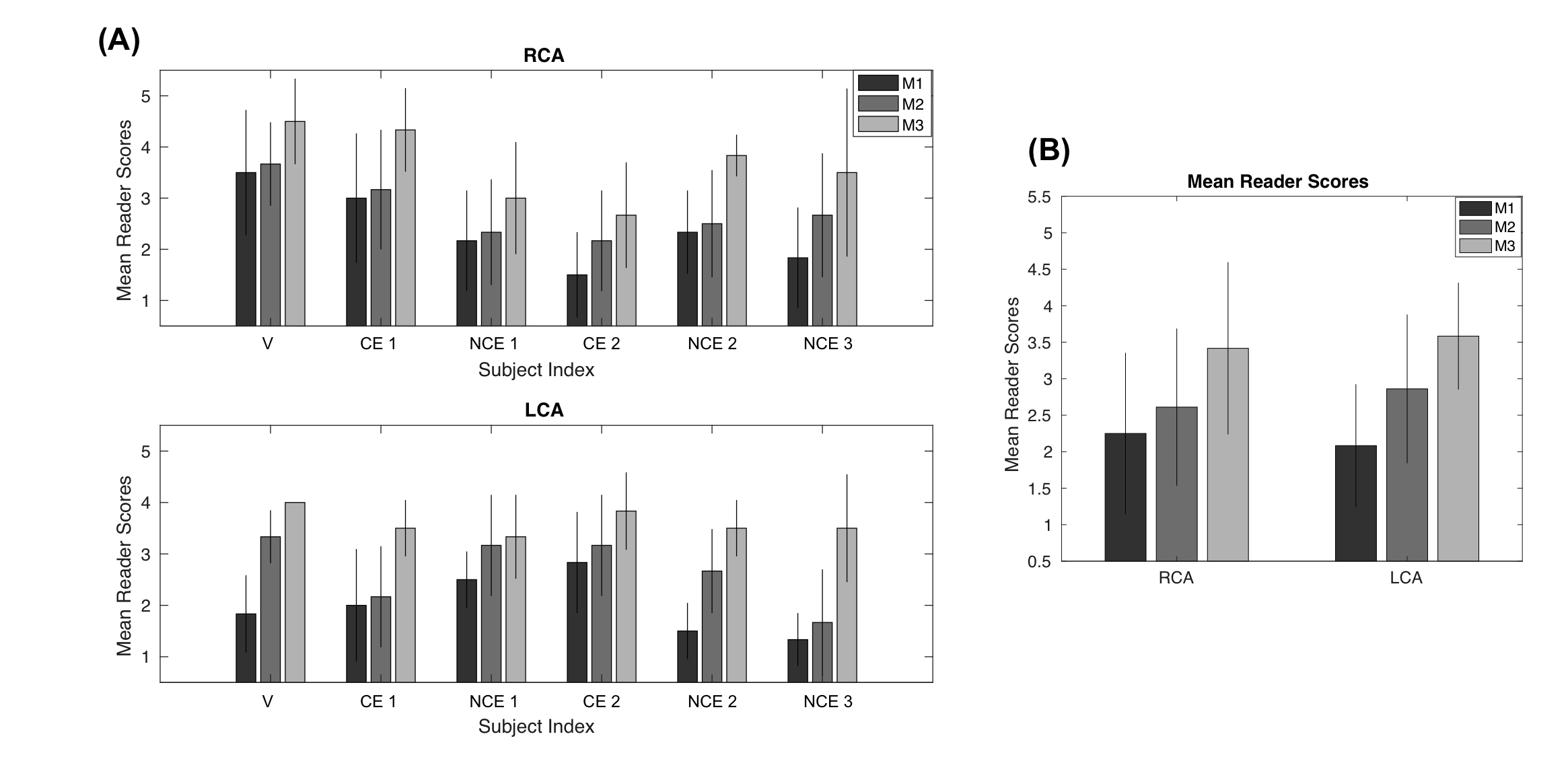}
  \caption[]
    {(A) The average of both reader scores for the LCA and RCA across all six studies (one volunteer (V), two contrast-enhanced (CE) patients, three non-contrast-enhanced (NCE) patients) and all three motion correction techniques (M1 = L\textsubscript{1}-ESPIRiT without motion correction, M2 = L\textsubscript{1}-ESPIRiT with translational correction, M3 = proposed motion-resolved reconstruction framework with intraphase 3D translational correction). (B) The mean RCA and LCA reader scores for the three different correction approaches. The statistical significance of the results for reader scores was P $<$ 0.01 for both the RCA and LCA when using the two-tailed Student's t-test. 
    }
\end{figure}

\begin{figure}[h]
  \centering
        \includegraphics[width=\linewidth]{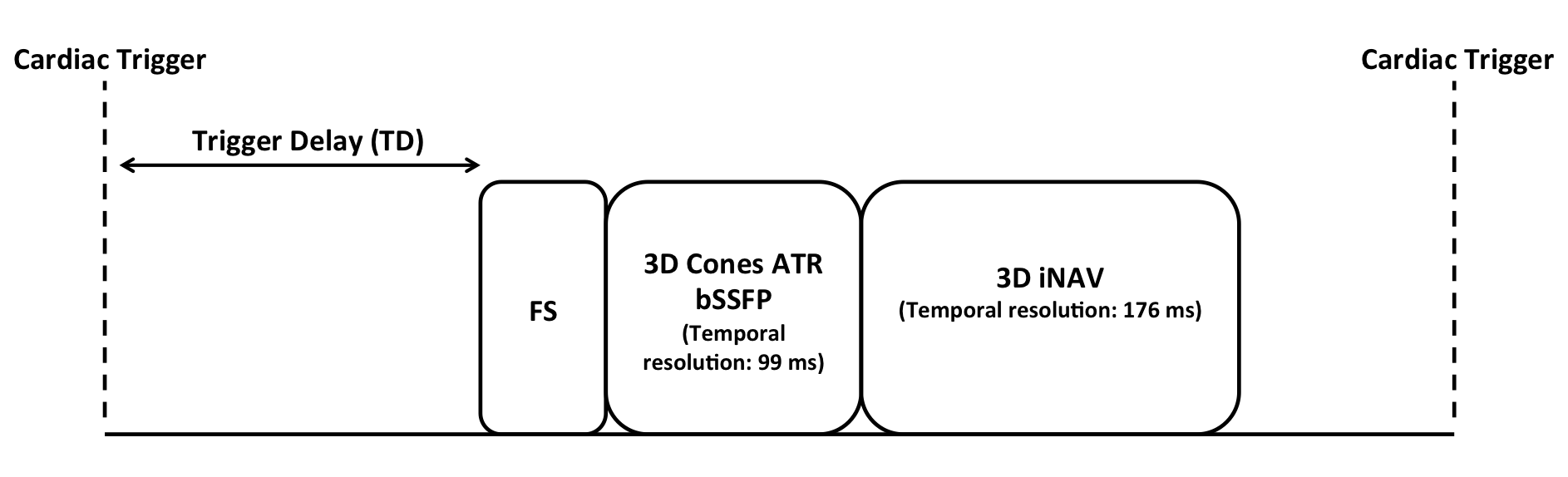}
  \caption*
    {Supporting Information Figure S1: Imaging data is collected with a cardiac-triggered sequence. A fat-saturation (FS) module is followed by a 3D cones sequence, where cones interleaves are acquired in groups of 18 with a temporal resolution of 99 ms during diastole. A 3D iNAV is also collected with a temporal resolution of 176 ms each heartbeat to monitor heart motion.
    }
\end{figure}

\begin{figure}[h]
  \centering
        \includegraphics[width=\linewidth]{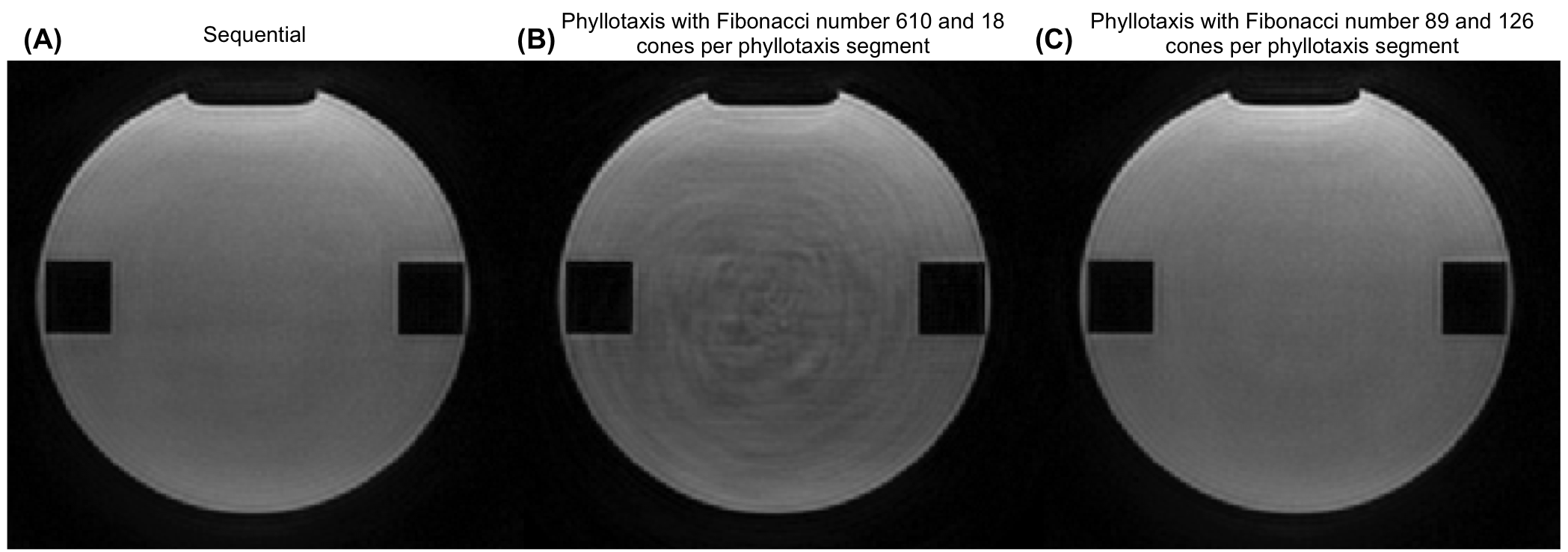}
  \caption*
    {Supporting Information Figure S2: Applying a phyllotaxis readout ordering strategy with 610 as the Fibonacci number (B) results in eddy current artifacts in a stationary 0.98 mm isotropic resolution phantom scan. Implementing cones reordering with 89 as the Fibonacci number (C) yields image quality comparable to that obtained from the standard sequential collection pattern (A). This corroborates the trends seen in \textit{in vivo} experimentation.
    }
\end{figure}

\begin{figure}[h]
  \centering
        \includegraphics[width=\linewidth]{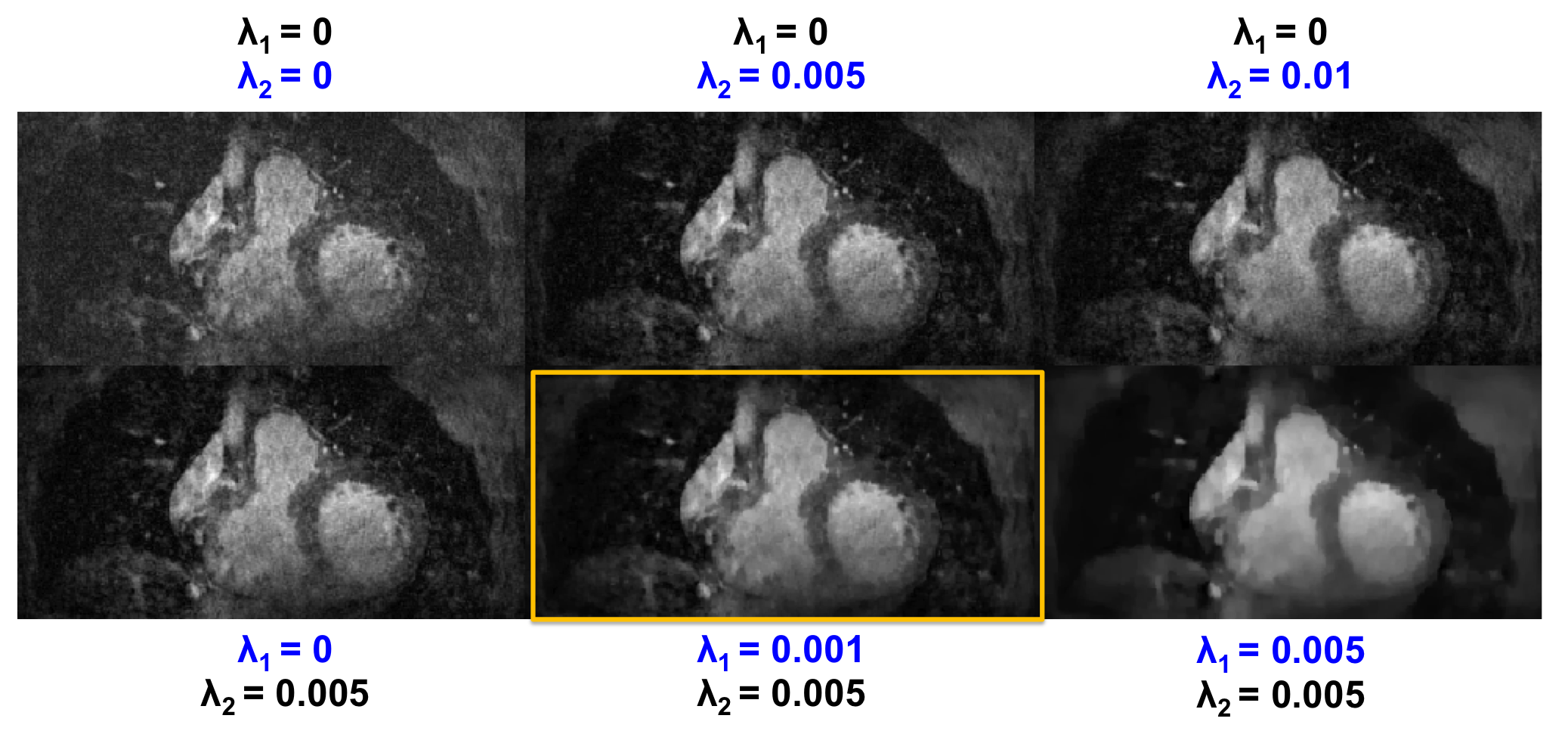}
  \caption*
    {Supporting Information Figure S3: Respiratory-resolved reconstructions of the end-expiration phase for different combinations of spatial ($\lambda_{1}$) and temporal ($\lambda_{2}$) regularization parameters. A large $\lambda_{2}$ introduces blurring across the motion states, while a very high value for $\lambda_{1}$ causes unwanted smoothening of the overall image. $\lambda_{1}$ = 0.001 and $\lambda_{2}$ = 0.005 provide a trade-off between noise reduction and blurring/smoothening effects.
    }
\end{figure}

\begin{figure}[h]
  \centering
        \includegraphics[width=\linewidth]{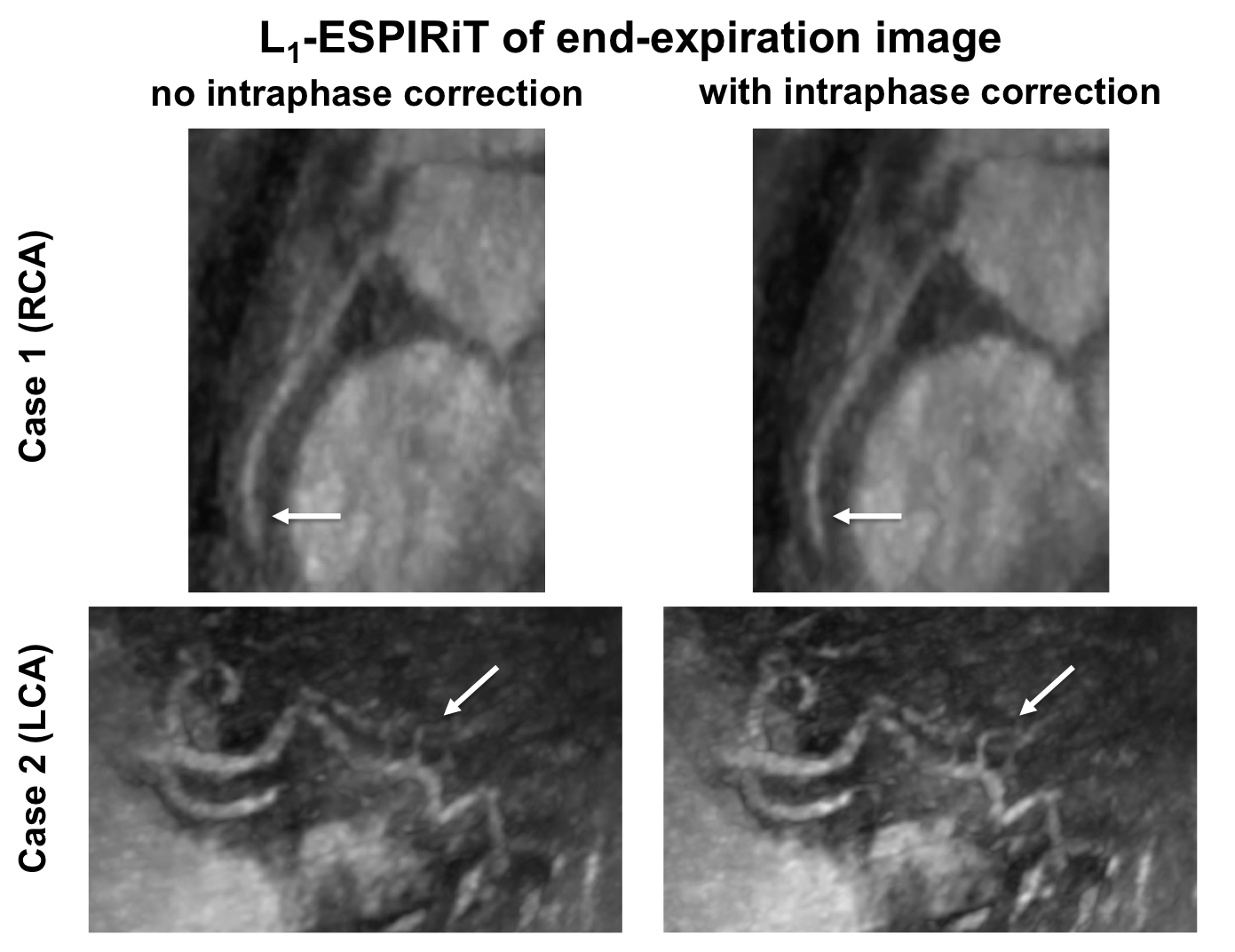}
  \caption*
    {Supporting Information Figure S4: 3D intraphase translational correction improves distal RCA sharpness in case 1 (non-contrast-enhanced). In case 2 (contrast-enhanced), a LCA branch is better depicted after accounting for the residual 3D translational motion in the end-expiration image. White arrows indicate regions of differences with intraphase correction. 
    }
\end{figure}

\begin{figure}[h]
  \centering
        \includegraphics[width=\linewidth]{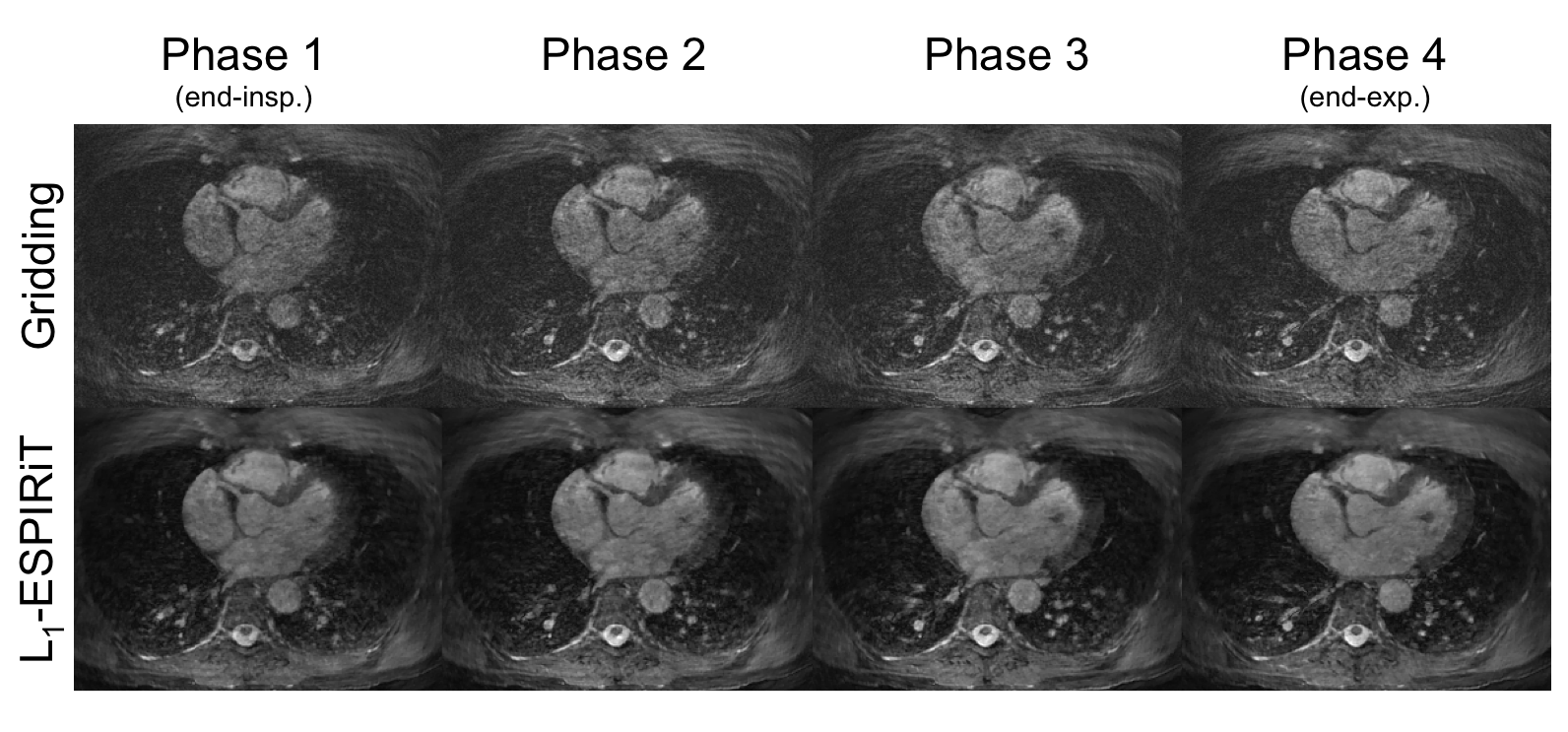}
  \caption*
    {Supporting Information Figure S5: Corresponding gridding and L\textsubscript{1}-ESPIRiT outcomes for the four respiratory phases from one subject study. The noise-like aliasing artifacts resulting from the undersampling in each phase are well-suited for compressed sensing reconstructions. The displacement of the coronary vessel can be seen across the different respiratory motion states.
    }
\end{figure}


%% file: main.bbl
\begin{thebibliography}{10}

\bibitem{rochitte2013computed}
Rochitte~CE, George~RT, Chen~MY, ArbabZadeh~A, Dewey~M, Miller~JM, Niinuma~H,
  Yoshioka~K, Kitagawa~K, Nakamori~S et~al.
\newblock Computed tomography angiography and perfusion to assess coronary
  artery stenosis causing perfusion defects by single photon emission computed
  tomography: the core320 study.
\newblock European Heart Journal 2013; 35:1120--1130.

\bibitem{hoffmann2006coronary}
Hoffmann~U, Ferencik~M, Cury~RC, Pena~AJ.
\newblock Coronary ct angiography.
\newblock Journal of Nuclear Medicine 2006; 47:797--806.

\bibitem{sakuma2006detection}
Sakuma~H, Ichikawa~Y, Chino~S, Hirano~T, Makino~K, Takeda~K.
\newblock Detection of coronary artery stenosis with whole-heart coronary
  magnetic resonance angiography.
\newblock Journal of the American College of Cardiology 2006; 48:1946--1950.

\bibitem{yang2009contrast}
Yang~Q, Li~K, Liu~X, Bi~X, Liu~Z, An~J, Zhang~A, Jerecic~R, Li~D.
\newblock Contrast-enhanced whole-heart coronary magnetic resonance angiography
  at 3.0-t: a comparative study with x-ray angiography in a single center.
\newblock Journal of The American College of Cardiology 2009; 54:69--76.

\bibitem{kato2010assessment}
Kato~S, Kitagawa~K, Ishida~N, Ishida~M, Nagata~M, Ichikawa~Y, Katahira~K,
  Matsumoto~Y, Seo~K, Ochiai~R et~al.
\newblock Assessment of coronary artery disease using magnetic resonance
  coronary angiography: a national multicenter trial.
\newblock Journal of the American College of Cardiology 2010; 56:983--991.

\bibitem{piccini2014respiratory}
Piccini~D, Monney~P, Sierro~C, Coppo~S, Bonanno~G, VanHeeswijk~RB, Chaptinel~J,
  Vincenti~G, DeBlois~J, Koestner~SC et~al.
\newblock Respiratory self-navigated postcontrast whole-heart coronary mr
  angiography: initial experience in patients.
\newblock Radiology 2014; 270:378--386.

\bibitem{prieto2015highly}
Prieto~C, Doneva~M, Usman~M, Henningsson~M, Greil~G, Schaeffter~T, Botnar~RM.
\newblock Highly efficient respiratory motion compensated free-breathing
  coronary mra using golden-step cartesian acquisition.
\newblock Journal of Magnetic Resonance Imaging 2015; 41:738--746.

\bibitem{gharib2012feasibility}
Gharib~AM, AbdElmoniem~KZ, Ho~VB, F{\"o}di~E, Herzka~DA, Ohayon~J, Stuber~M,
  Pettigrew~RI.
\newblock The feasibility of 350 $\mu$m spatial resolution coronary magnetic
  resonance angiography at 3 t in humans.
\newblock Investigative Radiology 2012; 47:339--345.

\bibitem{pang2015accelerated}
Pang~J, Sharif~B, Arsanjani~R, Bi~X, Fan~Z, Yang~Q, Li~K, Berman~DS, Li~D.
\newblock Accelerated whole-heart coronary mra using motion-corrected
  sensitivity encoding with three-dimensional projection reconstruction.
\newblock Magnetic Resonance in Medicine 2015; 73:284--291.

\bibitem{akccakaya2014accelerated}
Ak{\c{c}}akaya~M, Basha~TA, Chan~RH, Manning~WJ, Nezafat~R.
\newblock Accelerated isotropic sub-millimeter whole-heart coronary mri:
  compressed sensing versus parallel imaging.
\newblock Magnetic Resonance in Medicine 2014; 71:815--822.

\bibitem{addy2015high}
Addy~NO, Ingle~RR, Wu~HH, Hu~BS, Nishimura~DG.
\newblock High-resolution variable-density 3d cones coronary mra.
\newblock Magnetic Resonance in Medicine 2015; 74:614--621.

\bibitem{bustin2019five}
Bustin~A, Ginami~G, Cruz~G, Correia~T, Ismail~TF, Rashid~I, Neji~R, Botnar~RM,
  Prieto~C.
\newblock Five-minute whole-heart coronary mra with sub-millimeter isotropic
  resolution, 100\% respiratory scan efficiency, and 3d-prost reconstruction.
\newblock Magnetic Resonance in Medicine 2019; 81:102--115.

\bibitem{nagel1999optimization}
Nagel~E, Bornstedt~A, Schnackenburg~B, Hug~J, Oswald~H, Fleck~E.
\newblock Optimization of realtime adaptive navigator correction for 3d
  magnetic resonance coronary angiography.
\newblock Magnetic Resonance in Medicine 1999; 42:408--411.

\bibitem{nehrke2001free}
Nehrke~K, Bornert~P, Manke~D, Bock~JC.
\newblock Free-breathing cardiac mr imaging: study of implications of
  respiratory motion: initial results.
\newblock Radiology 2001; 220:810--815.

\bibitem{moghari2012subject}
Moghari~MH, Hu~P, Kissinger~KV, Goddu~B, Goepfert~L, Ngo~L, Manning~WJ,
  Nezafat~R.
\newblock Subject-specific estimation of respiratory navigator tracking factor
  for free-breathing cardiovascular mr.
\newblock Magnetic Resonance in Medicine 2012; 67:1665--1672.

\bibitem{manke2002respiratory}
Manke~D, Nehrke~K, B{\"o}rnert~P, R{\"o}sch~P, D{\"o}ssel~O.
\newblock Respiratory motion in coronary magnetic resonance angiography: a
  comparison of different motion models.
\newblock Journal of Magnetic Resonance Imaging 2002; 15:661--671.

\bibitem{shechter2004respiratory}
Shechter~G, Ozturk~C, Resar~JR, McVeigh~ER.
\newblock Respiratory motion of the heart from free breathing coronary
  angiograms.
\newblock IEEE Transactions on Medical Imaging 2004; 23:1046--1056.

\bibitem{ingle2014nonrigid}
Ingle~RR, Wu~HH, Addy~NO, Cheng~JY, Yang~PC, Hu~BS, Nishimura~DG.
\newblock Nonrigid autofocus motion correction for coronary mr angiography with
  a 3d cones trajectory.
\newblock Magnetic Resonance in Medicine 2014; 72:347--361.

\bibitem{addy20173d}
Addy~NO, Ingle~RR, Luo~J, Baron~CA, Yang~PC, Hu~BS, Nishimura~DG.
\newblock 3d image-based navigators for coronary mr angiography.
\newblock Magnetic Resonance in Medicine 2017; 77:1874--1883.

\bibitem{luo2017nonrigid}
Luo~J, Addy~NO, Ingle~RR, Baron~CA, Cheng~JY, Hu~BS, Nishimura~DG.
\newblock Nonrigid motion correction with 3d image-based navigators for
  coronary mr angiography.
\newblock Magnetic Resonance in Medicine 2017; 77:1884--1893.

\bibitem{cruz2017highly}
Cruz~G, Atkinson~D, Henningsson~M, Botnar~RM, Prieto~C.
\newblock Highly efficient nonrigid motion-corrected 3d whole-heart coronary
  vessel wall imaging.
\newblock Magnetic Resonance in Medicine 2017; 77:1894--1908.

\bibitem{correia2018accelerated}
Correia~T, Cruz~G, Schneider~T, Botnar~RM, Prieto~C.
\newblock Accelerated nonrigid motion-compensated isotropic 3d coronary mr
  angiography.
\newblock Medical Physics 2018; 45:214--222.

\bibitem{feng2016xd}
Feng~L, Axel~L, Chandarana~H, Block~KT, Sodickson~DK, Otazo~R.
\newblock Xd-grasp: golden-angle radial mri with reconstruction of extra
  motion-state dimensions using compressed sensing.
\newblock Magnetic Resonance in Medicine 2016; 75:775--788.

\bibitem{piccini2017four}
Piccini~D, Feng~L, Bonanno~G, Coppo~S, Yerly~J, Lim~RP, Schwitter~J,
  Sodickson~DK, Otazo~R, Stuber~M.
\newblock Four-dimensional respiratory motion-resolved whole heart coronary mr
  angiography.
\newblock Magnetic Resonance in Medicine 2017; 77:1473--1484.

\bibitem{feng20185d}
Feng~L, Coppo~S, Piccini~D, Yerly~J, Lim~RP, Masci~PG, Stuber~M, Sodickson~DK,
  Otazo~R.
\newblock 5d whole-heart sparse mri.
\newblock Magnetic Resonance in Medicine 2018; 79:826--838.

\bibitem{correia2018optimized}
Correia~T, Ginami~G, Cruz~G, Neji~R, Rashid~I, Botnar~RM, Prieto~C.
\newblock Optimized respiratory-resolved motion-compensated 3d cartesian
  coronary mr angiography.
\newblock Magnetic Resonance in Medicine 2018; 80:2618--2629.

\bibitem{gurney2006design}
Gurney~PT, Hargreaves~BA, Nishimura~DG.
\newblock Design and analysis of a practical 3d cones trajectory.
\newblock Magnetic Resonance in Medicine 2006; 55:575--582.

\bibitem{wu2013free}
Wu~HH, Gurney~PT, Hu~BS, Nishimura~DG, McConnell~MV.
\newblock Free-breathing multiphase whole-heart coronary mr angiography using
  image-based navigators and three-dimensional cones imaging.
\newblock Magnetic Resonance in Medicine 2013; 69:1083--1093.

\bibitem{malave2019whole}
Malav{\'e}~MO, Baron~CA, Addy~NO, Cheng~JY, Yang~PC, Hu~BS, Nishimura~DG.
\newblock Whole-heart coronary mr angiography using a 3d cones phyllotaxis
  trajectory.
\newblock Magnetic Resonance in Medicine 2019; 81:1092--1103.

\bibitem{piccini2011spiral}
Piccini~D, Littmann~A, NiellesVallespin~S, Zenge~MO.
\newblock Spiral phyllotaxis: the natural way to construct a 3d radial
  trajectory in mri.
\newblock Magnetic Resonance in Medicine 2011; 66:1049--1056.

\bibitem{bieri2005analysis}
Bieri~O, Markl~M, Scheffler~K.
\newblock Analysis and compensation of eddy currents in balanced ssfp.
\newblock Magnetic Resonance in Medicine 2005; 54:129--137.

\bibitem{zucker2018free}
Zucker~EJ, Cheng~JY, Haldipur~A, Carl~M, Vasanawala~SS.
\newblock Free-breathing pediatric chest mri: performance of self-navigated
  golden-angle ordered conical ultrashort echo time acquisition.
\newblock Journal of Magnetic Resonance Imaging 2018; 47:200--209.

\bibitem{rohde2004comprehensive}
Rohde~GK, Barnett~A, Basser~P, Marenco~S, Pierpaoli~C.
\newblock Comprehensive approach for correction of motion and distortion in
  diffusion-weighted mri.
\newblock Magnetic Resonance in Medicine 2004; 51:103--114.

\bibitem{uecker2014espirit}
Uecker~M, Lai~P, Murphy~MJ, Virtue~P, Elad~M, Pauly~JM, Vasanawala~SS,
  Lustig~M.
\newblock Espirit—an eigenvalue approach to autocalibrating parallel mri:
  where sense meets grappa.
\newblock Magnetic Resonance in Medicine 2014; 71:990--1001.

\bibitem{batchelor2005matrix}
Batchelor~P, Atkinson~D, Irarrazaval~P, Hill~D, Hajnal~J, Larkman~D.
\newblock Matrix description of general motion correction applied to multishot
  images.
\newblock Magnetic Resonance in Medicine 2005; 54:1273--1280.

\bibitem{odille2008generalized}
Odille~F, Vuissoz~PA, Marie~PY, Felblinger~J.
\newblock Generalized reconstruction by inversion of coupled systems (grics)
  applied to free-breathing mri.
\newblock Magnetic Resonance in Medicine 2008; 60:146--157.

\bibitem{horn1981determining}
Horn~BK, Schunck~BG.
\newblock Determining optical flow.
\newblock Artificial Intelligence 1981; 17:185--203.

\bibitem{rueckert1999nonrigid}
Rueckert~D, Sonoda~LI, Hayes~C, Hill~DL, Leach~MO, Hawkes~DJ.
\newblock Nonrigid registration using free-form deformations: application to
  breast mr images.
\newblock IEEE Transactions on Medical Imaging 1999; 18:712--721.

\bibitem{sigpy}
Ong~F, Lustig~M.
\newblock Sigpy: a python package for high performance iterative
  reconstruction.
\newblock International Society for Magnetic Resonance in Medicine 2019; p.
  4819.

\bibitem{baron2018rapid}
Baron~CA, Dwork~N, Pauly~JM, Nishimura~DG.
\newblock Rapid compressed sensing reconstruction of 3d non-cartesian mri.
\newblock Magnetic Resonance in Medicine 2018; 79:2685--2692.

\end{thebibliography}
